# Low-power In-pixel Computing with Current-modulated Switched Capacitors

David Zhang, Gooitzen van der Wal, Saurabh Farkya, Thomas Senko, Aswin Raghavan,
Michael Isnardi, Michael Piacentino

Center for Vision Technologies, SRI International, Princeton, NJ
{David.Zhang, Gooitzen.Vanderwal, Saurabh.Farkya, Thomas.Senko, Aswin.Raghavan, Michael.Isnardi, Michael.Piacentino}@sri.com

## Abstract

*We present a scalable in-pixel processing architecture that can reduce the data throughput by 10X and consume less than 30 mW per megapixel at the imager frontend. Unlike the* state-of-the-art (*SOA*) *analog process-in-pixel (PIP) that modulates the exposure time of photosensors when performing matrix-vector multiplications, we use switched capacitors and pulse width modulation (PWM). This non-destructive approach decouples the sensor exposure and computing, providing processing parallelism and high data fidelity. Our design minimizes the computational complexity and chip density by leveraging the patch-based feature extraction that can perform as well as the CNN. We further reduce data using partial observation of the attended objects, which performs closely to the full frame observations. We have been studying the reduction of output features as a function of accuracy, chip density and power consumption from a transformer-based backend model for object classification and detection.*

## 1. Introduction

Today's AI processing of video presents a unique difficult problem in bringing AI to the sensor edge. The high resolution, high dynamic range, and high frame rates generate significantly more data in real time than other edge sensing modalities. To accommodate this large data stream, SOA DNNs for computer vision tasks like multiple object tracking (MOT) employ one or more convolutional NNs (CNNs) that are 10-100 layers deep. Methods that only reduce the complexity of operations (e.g., quantization) are *not* applicable as they do not reduce the high data bandwidth. The large data throughput and overwhelming complexity of NNs has separated data capture from data analysis that is carried out by backend computers and cloud servers. The large data transfer is expensive, slow and has long latency. Solutions compromising to power, size and latency constraints have been proposed at the mobile edge by reducing the size of NNs as processing units are moved closer to the sensor. In addition to the degradation of accuracy, the reduction in distance does not solve data complexity and high-power problem since processing is still performed after data movement.

To move the processing into the sensor pixel before the data movement is an effective way to reduce power and bandwidth at the sensor edge. To achieve this, we concentrated on the following 3 concepts:

**Saliency pixels** - Identifying salient pixels (or features) holds promise if saliency can be extracted efficiently (in-pixel) and streamed for downstream processing. It is evident that an image is spatially redundant, and a temporal video is spatiotemporally redundant to the task at hand [1]. Instead of relying on the entire image or video, Flexgrid2vec (2021) [2] forms a graph of patches around key points, uses a GNN to extract vector representations, and obtains SOA accuracy with <10% pixels used. PixelNet [3] randomly samples 4% of pixels, extracts features from a CNN, and reports SOA performance.

**Patch based processing** - Image patches, which are treated as tokens in an NLP application [4], has been proven to attain results comparable to the SoA full-image convolutional networks in classification, detection, and tracking [5-13].

**Shifted attention** - Saccade is a quick movement of the eyes between two or more phases of fixation in the same direction. This biological attention-shift mechanism [1,14--16] focuses on a few attentive object features to be processed quickly while maintaining overall saliency and spatiotemporal prediction to be effectively performed on one or more objects in high-resolution high frame rate vision sensors.

Thus, we propose an in-pixel processing (IP2) analog design that computes salient patch features before ADC. In this paper, we do not focus on the study of saccadic patch prediction and performance of the end-to-end deep computer vison models [17], rather we design the in-pixel circuits leveraging these advanced features. Our main contribution is listed as follows:

1. Designed a patch-based linear projection model (patch size matrix multiply) in the analog circuits before ADC. We are in the process of studying the performance

This research was, in part, funded by the U.S. Government. The views and conclusions contained in this document are those of the authors and should not be interpreted as representing the official policies, either expressed or implied, of the U.S. Government.

in terms of data reduction and accuracy in consideration of the technology node, chip density and stack options.

2. Proposed a non-destructive approach that decouples the sensor exposure and computes the matrix-vector multiplications using switched capacitors and pulse width modulation (PWM).

3. Support dimensionality reduction in the analog domain, assuming 25% patches are active per frame.

4. Developed and simulated a scalable in-pixel processing architecture that can reduce the data dimensionality by 10X and consume less than 30 mW per megapixel at the imager frontend, including the ADC and DAC power.

## 2. In-pixel processing design and circuit implementation

Performing in-pixel and near-pixel analog processing, including neuromorphic processing [18], has been a goal for many years to reduce system power. With the advent of CNNs the goal is not only to reduce the front-end power, but also to reduce the power of the remaining processing by reducing the data rate at the output of the sensor. In a typical recent design, the in-pixel convolution / matrix multiply is achieved by controlling the exposure time as the weight followed by read-out to ADC [19,20]. For positive and negative weights this results in 8 ADCs per 3x3 convolution. This results in a very small in-pixel circuit but requires very short integration times and operation is influenced by sensor motion. Other approaches use near-pixel analog processing [21].

The approach in this paper emphasizes the need for full exposure range, while performing the analog processing in-pixel, in parallel on global shuttered data during the frame period, and a flexible size linear patch-based feature generation while reducing the output data rate of the sensor.

### 2.1. Non-destructive in-pixel circuits

The in-pixel frontend is based on available CMOS Continuous Interleaved Sampling (CIS) processes utilizing a pinned photo diode detector with in-pixel capacitor storage of clamp and sample signals. We propose to use switched capacitor techniques allowing fabrication in deep sub-micron processes and minimizing power consumption. The multiplication of input signals with programmed weights are realized by time modulation of capacitor charging current. The circuit implementation regarding the key operations in deep NN is briefly summarized below:

**Multiplication/division** – A photon signal *P* can be multiplied with or divided by a weight *w* programmed by pulse width modulation (PWM) to modulate the capacitor charging current.

**Quantized division** - The application of weights to the pixel correlated double sampling (CDS) voltage output is achieved by charging a capacitor to the CDS voltage output. Then additional capacitors are switched onto the first capacitor to divide down the voltage and produce a weighted voltage output.

**Weighted sum** - The *addition* of the weighted outputs of two pixels is realized by switches that connect the weighted voltages in series. The *Subtraction* operation is enabled by reversing the capacitor polarity before connecting in series or applying an inverse modulated charging current.

**ReLU activation** – A two-transistor (2T) method is used to convert pixel voltage into a current. This voltage controlled current source modulates the drain to source voltage of a FET biased in the linear region – either as a *ReLU* or a *Sigmoid* depending on the bias voltage.

**Pixel selection** - When a pixel is located in a deselected patch the photo diode needs to be cleared so the next read is not contaminated by previously collected remnant charge. This operation requires very low power. For a BSI (Back-side-illuminated) imager this is implemented using a lateral charge dump drain and control gate. Any unused transistors are powered down. The front-end source follower consumes no power unless the clamp storage capacitor is clocked.

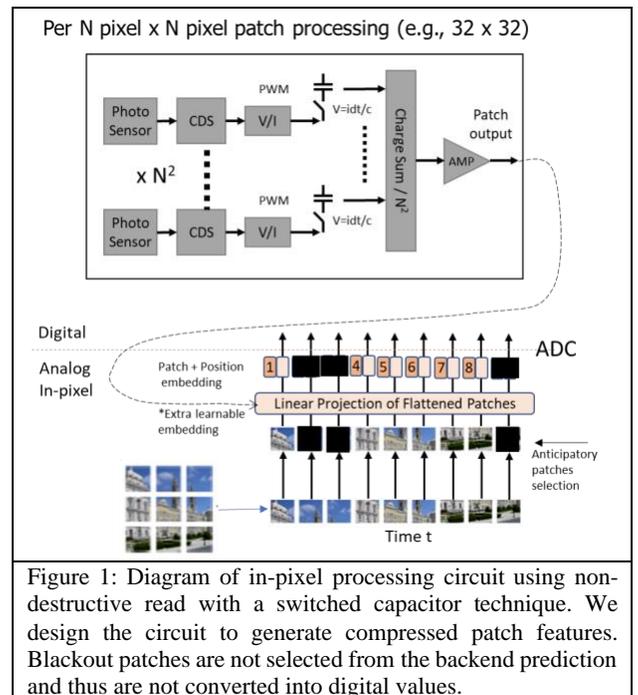

Figure 1: Diagram of in-pixel processing circuit using non-destructive read with a switched capacitor technique. We design the circuit to generate compressed patch features. Blackout patches are selected from the backend prediction and thus are not converted into digital values.

**In-pixel circuitry** - Figure 1 is our high-level in-pixel circuit design. All pixels are exposed at the same time and sampled and held with a CDS circuit. They are the input to computing the linear projection in the analog domain M times for an M-size analog 1-D vector per patch. The exposure time is in parallel with the processing providing global shutter implementation with a programmable exposure time up to the frame period.

The sampled values of each frame are the inputs to the analog vector multiplication using pulse-width modulation and programmable weights currents for each of the pixels with the result stored in a small capacitor for each pixel. The other side of the capacitors are connected to a single amplifier per patch to hold the summed value. The charge is distributed between all the capacitors computing the sum divided by the number of pixels in the patch. By using an amplifier, the summed charge is held constant (Section 2.1.2). This analog vector computation is performed M times to generate an M-size analog 1-D vector per patch.

The basic in-pixel linear projection circuit shown in Figure 2 uses a CDS circuit to sample and hold the exposed pixels for a full frame. Then a weighted sum is computed over an $N^2$ pixel patch M times for an M 1-D vector per patch. For every vector after a reset, for each pixel a capacitor is charged with a current that represents the pixel intensity converted to a pulse width, charging the pixel capacitor with weight voltage controlled current, generating a charge that is the multiplication of the pixel and the programmed weight. For negative weights, a negative current is used. Then a switch is used to connect all the capacitors in the patch to the negative input of the amplifier, with the positive input connected to a voltage reference $V_R$. The analog output for vector $v$ with weights $W_{i,v}$ and pixels $P_i$ for $i = 1..N^2$ then represents:

$$Out_v = V_R + \sum_{N^2} (W_{i,v} \cdot P_i)/N^2$$

For the baseline design the patch outputs are read-out to an ADC at the edge. This $Out_v$ is the readout to an ADC with $V_R - b$ subtracted in the digital domain to compute:

$$Dig\_Out_v = b + \sum_{N^2} (W_{i,v} \cdot P_i)/N^2$$

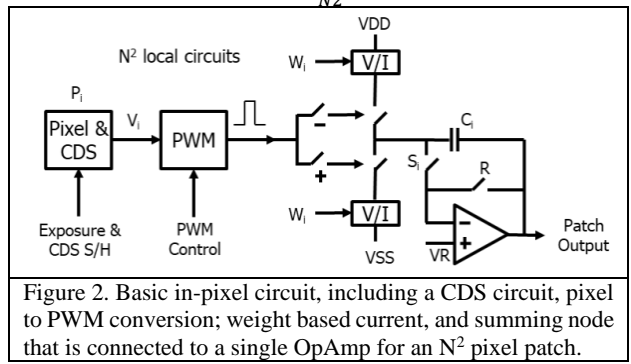

Figure 2. Basic in-pixel circuit, including a CDS circuit, pixel to PWM conversion; weight based current, and summing node that is connected to a single OpAmp for an $N^2$ pixel patch.

Only the outputs of a selected set of salient patches (e.g., < 25%) are converted to the digital domain and used for the next computing layer, thus saving power, output dimension and bandwidth.

### 2.1.1 Programmable patch size

The linear projection circuit described is limited to non-overlapping NxN size patches with one OpAmp per NxN patch per output vector. Our baseline design is for an 8x8 minimum patch size that can be offset by 4 pixels in either direction per vector. Multiple patches can be connected for any of the vector computations for patch sizes of 8,16,24 or 32 pixels in horizontal and vertical directions (e.g., 8 x 32, 24 x 16, etc.). We selected 8x8 minimum patches to limit the number of in-pixel OpAmps, but other sizes are certainly possible. The pixels can either be monochrome or have a color pattern such as Bayer.

### 2.1.2 Switch leakage and amplifier for compensation

The proposed circuit using switched capacitors as weights to multiply source signals and sum into an output is shown in Figure 1. We simulated 768 capacitors charged with 1V summed with 768 capacitors charged with 0V. The expected result is 0.5V. In a 65nm process the thin oxide MOSFET switches leak and the *passive* capacitive summer reduces by 10% in under 10 µsec. To fix this problem, we are summing the capacitors in the feedback loop of an amplifier, which stabilizes the output as the amplifier compensates for the leakage. The amplifiers can be removed when using a lower leakage technology (i.e., 22 nm FDSOI transistors), and/or process with reduced accuracy.

### 2.1.3 In-pixel circuit size

The initial conservative component size estimates are listed in Table 1 assuming 65nm technology with 8 µm pixels and 30 fF capacitors, one OpAmp per patch and an estimate for wiring, the in-pixel circuit size is 22 µm pitch. The estimate power is < 60 mW for a 2 Mpix sensor @ 30 Hz processing and capture rate excluding the digital interface. The majority of the power is for the ADC conversion, and assumes that only 25% of the patches generate an output every frame.

**Table 1**. In-pixel circuit size per pixel 65nm

| Pixel size | # | Size (µm²) | Total (µm²) | Occupancy |
|---|---|---|---|---|
| Photo Sensor | 1 | 64 | 64 | 13% |
| Cap 30 fF | 3 | 64 | 192 | 40% |
| Transistors | 41 | 5 | 205 | 42% |
| Wiring | 1 | 16 | 16 | 3% |
| Margin | 1 | 8 | 8 | 2% |
| Total | | | 485 | 100% |
| Pixel pitch | | | 22.0 µm | |

The size of the capacitors for the CDS and accumulation and the size of the pixel dominate and can be significantly reduced while reducing the accuracy of the computations. Our current simulations indicate a 6-bit accuracy in the in-pixel processing. Other research has suggested that precision can be reduced down to 1-bit weights and intermediate results [22,23].

A reduction in technology (i.e., 22nm FDX) can provide reduction in pixel size and using a two-stacked die would reduce the pixel design further. A preferred design would be a backside illuminated sensor stacked to a second die with additional analog processing and ADC on the edge. Interconnect between these two dies, which may use different technologies, is between the top metal layers of the two dies, so through silicon vias are not required.

### 2.1.4 Processing rate and throughput

The processing rate is a function of many parameters, including the size of the patches, the number of vectors computer per patch for the full frame, the number of ADCs and weight DACs, etc. By increasing the number of weight DACs and weight wires per pixel column in Figure 3(a), the throughput increases. Figure 3(b) shows the throughput of various image formats vs feature output sizes. This means that there will be 1,2,4, or 8 weight voltage lines per pixel column. The figure indicates that we can achieve 100 Mpix/s for a 1080P imager with 2 weight lines for 400 vectors per 32 x 32 patch at a frame rate of 90 Hz with a data dimension reduction of 10x. The effective data dimensionality reduction is 30x when comparing to RGB data, since the Bayer to RGB interpolation is included in the linear projection. For 8x8 patches, generating 192 vectors/patch the frame rate is > 30 Hz.

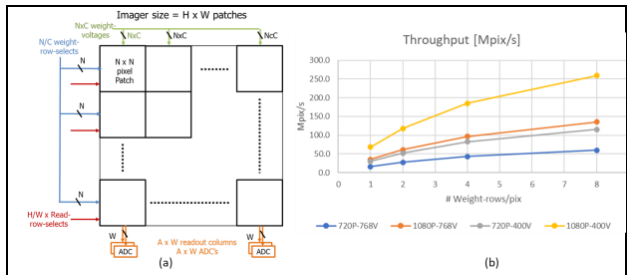

Figure 3. (a) Programing the weights for multiple rows per patch. (b) Rate increase for C=1,2,4, and 8 rows per weight programming & compute cycles for 720P and 1080P sensor and for 400 and 768 vectors / 32x32 patch.

### 2.1.5 Sensor Bayer pattern and anti-aliasing

The algorithm Simulation had an input image size of NxN x 3 color channels. Howeve7, the hardware (HW) sensor will produce a raw, mosaiced Bayer-pattern image of size N x N pixels. No demosaicing will be performed in HW, as the Bayer pattern image would be used directly as input without interpolating to 3-channel RGB.

The linear mapping of vectorized input RGB pixels, represented in simulation by a trained matrix **A**, was transformed to a new matrix **A'** by striking out those columns of **A** that have no corresponding elements between the simulation and HW RGB vectors.

The in-pixel circuit assumes that the pixels, each with either an R, G, or B color filter to implement the Bayer format are far apart. To minimize aliasing effects with up to 22 mm pixel pitch we assume that micro lens technology is used to focus the image data to the photo diodes, providing fill factor near unity.

In simulation, we applied antialiasing filters that represent the combined optics of lenslets and the main camera lens. We applied Gaussian spatial lowpass filters having magnitude responses with 0.5 and 0.25 Nyquist cutoff (-3dB) frequencies. We found that the training accuracy is virtually unaffected, even with the 0.25 Nyquist cutoff, indicating that slightly defocusing the camera lens will act as a good antialiasing filter and will not affect the accuracy of the results. This also means that the in-pixel processing sensor may provide good results with a sensor at ½ the resolution of the example full resolution RGB data for the tasks required for this program. For example, if the original trained data is 1920 x 1080 RGB, the current in-pixel processing design may be able to achieve accurate results with a 960 x 540 Bayer pixel array.

## 3. Discussions and future work

We carried out an early study on low-power in-pixel processing study using non-destructive in-pixel circuits on the selective salient patches to reduce power, feature size and bandwidth while maintaining high throughput and model accuracy. Further study include reducing noise and distortion in analog circuit, analyzing scaleup design of multiple patch-based layers when running end-to-end simulation models such as Swin-transformers [6], GAT [5], and DeTR [24]. We will leverage the codesign study and performance analysis reported from [17].

Our analog design can be extended as suggested in Figure 4, which shows a self-attention [5] circuit approximates the quadratic relations in the attention by mapping the attention coefficients from each neighborhood $i$ into a quantized (QTH) power-of-2 weight.

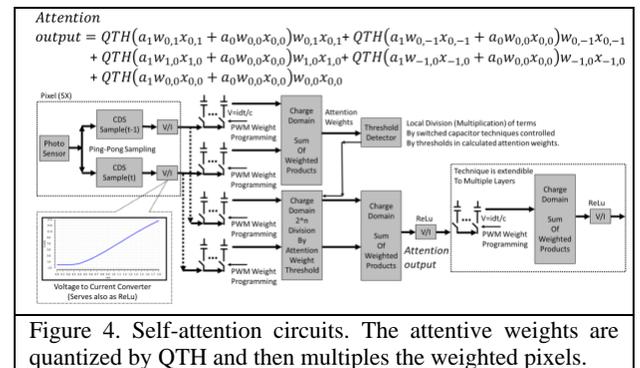

Figure 4. Self-attention circuits. The attentive weights are quantized by QTH and then multiples the weighted pixels.

In this design, the value representing an element in the patch feature can be stored in a second layer of patch processing modules that are similar to the linear projection module without the photon sensor.

### Acknowledgement
We offer special thanks to Dr. Mason, program manager